\documentclass[fleqn,10pt]{wlscirep}
\usepackage[utf8]{inputenc}
\usepackage[T1]{fontenc}

\usepackage{lineno}

\usepackage{booktabs}
\usepackage{multirow}
\usepackage{longtable}
\usepackage{multicol}
\usepackage{threeparttable}
\usepackage{tabularx}
\usepackage{array}

\newcolumntype{Y}{>{\raggedright\arraybackslash}X}

\makeatletter
\newcommand{\beginsupplementaryinformation}{%
  \clearpage
  \newgeometry{left=3cm,right=3cm,top=3cm,bottom=3cm}%
  \pagestyle{plain}%
  \setlength{\parindent}{15pt}%
  \setlength{\parskip}{0pt}%
  \renewcommand{\rmdefault}{ptm}%
  \normalfont
  \fontsize{11}{13.6}\selectfont
  \captionsetup{labelfont=bf,textfont=normalfont,labelsep=period,justification=justified,singlelinecheck=false}%
  \setcounter{figure}{0}%
  \setcounter{table}{0}%
  \renewcommand{\thefigure}{S\arabic{figure}}%
  \renewcommand{\thetable}{S\arabic{table}}%
  \setcounter{section}{0}%
  \setcounter{subsection}{0}%
  \setcounter{subsubsection}{0}%
  \titleformat{\section}{\normalfont\Large\bfseries}{\thesection}{0.75em}{##1}%
  \titleformat{name=\section,numberless}{\normalfont\Large\bfseries}{}{0em}{##1}%
  \titleformat{\subsection}{\normalfont\large\bfseries}{\thesubsection}{0.75em}{##1}%
  \titleformat{\subsubsection}{\normalfont\normalsize\itshape}{\thesubsubsection}{0.75em}{##1}%
  \titleformat{\paragraph}[runin]{\normalfont\normalsize\bfseries}{}{0pt}{##1}%
  \titlespacing*{\section}{0pt}{3.5ex plus 1ex minus .2ex}{2.3ex plus .2ex}%
  \titlespacing*{\subsection}{0pt}{3.25ex plus 1ex minus .2ex}{1.5ex plus .2ex}%
  \titlespacing*{\subsubsection}{0pt}{3.25ex plus 1ex minus .2ex}{1.5ex plus .2ex}%
  \titlespacing*{\paragraph}{0pt}{1ex}{1em}%
}

\newcommand{\supplementarytableofcontents}{%
  {\centering\large\bfseries Supplementary Notes\par}%
  \vspace{0.75\baselineskip}%
  \@starttoc{stc}%
}
\makeatother

\title{%
The benefits and biases of seeing the world's cities through marathons
}

\author[a,*]{Andrew Renninger}

\affil[a]{School of Geographical \& Earth Sciences, University of Glasgow}

\affil[*]{Corresponding author: Andrew Renninger (E-mail: andrew.renninger@glasgow.ac.uk)}

\begin{abstract}
Marathons are now common ways of seeing cities, yet little is known about how representative their routes are. Using 311 marathon routes across five continents, we compare landmarks and amenities along the course with those elsewhere in the same city, finding that museums are 15.7 times denser near the route and that the median city has about 8.5 times more luxury brands near the route than elsewhere in the city. These patterns persist under perturbed routes with the same start and finish lines: monuments and landmarks, in particular, are more prevalent on the race course than on similar alternative routes, suggesting that marathons function as intentionally selective urban portraits.
\end{abstract}

\begin{document}

\flushbottom
\maketitle

% \begin{linenumbers}

\section*{Introduction}
Running has grown since the pandemic, when many people took to the sport as a way to stay active while confined to home \cite{england2020running}. Strava reported a 9\% increase in the number of marathons, ultramarathons and century rides logged in 2024 \cite{strava2024yearinsport}, and more than 1 million people entered the lottery for the 2025 London Marathon \cite{nyt2025marathontours}. Running is also social and exploratory: running clubs are growing \cite{strava2024yearinsport}, they can provide important scaffolding for social life \cite{herrick2023running}, and travel around sport and exercise has grown, with the share of athletes logging activities outside their home country rising by 101\% between 2021 and 2022 \cite{strava2022yearinsport}. In 2025, 35\% of the runners in the Paris Marathon came from a different country to race \cite{paris2025marathon}. 

As more people see cities through running, what they see matters. Tourism research has long shown that destination image shapes tourist satisfaction and return intentions \cite{chi2008destination}, and that the image places project can differ from the one visitors actually perceive \cite{sun2021projected}. In marathon tourism specifically, race settings shape destination image and revisit intentions \cite{duan2024marathonimage}, while event zones can keep visitors concentrated in narrow parts of the host city \cite{duignan2023eventzones}. A broader literature on mobility makes a similar point: the places people traverse, not just the places where they live, shape exposure \cite{sharp2015activityspaces}. Exposure can also matter for attitudes and inference \cite{pettigrew2006contact, achard2025refugees, tropp2022youthcontact}. If marathon routes systematically privilege certain districts, they may shape not only what runners remember about a city, but what they infer about urban life itself.

In the following brief communication, we ask a narrow question with broad implications: what kind of city do marathons convey? Using 311 marathon routes across five continents, matched to data on landmarks, amenities, buildings, and population, we compare what lies along the course with what lies elsewhere in the same city. Marathons do run through dense central cities, but not all dense central city is the same. Across our sample, routes are disproportionately close to museums, attractions, luxury shops, premium hotels, and Apple Stores, while schools, grocers, and other everyday amenities are less favoured. These patterns remain even when we compare the observed route to plausible alternatives with the same start and finish lines.

\begin{figure*}[ht!]
\centering
\includegraphics[width=1\textwidth]{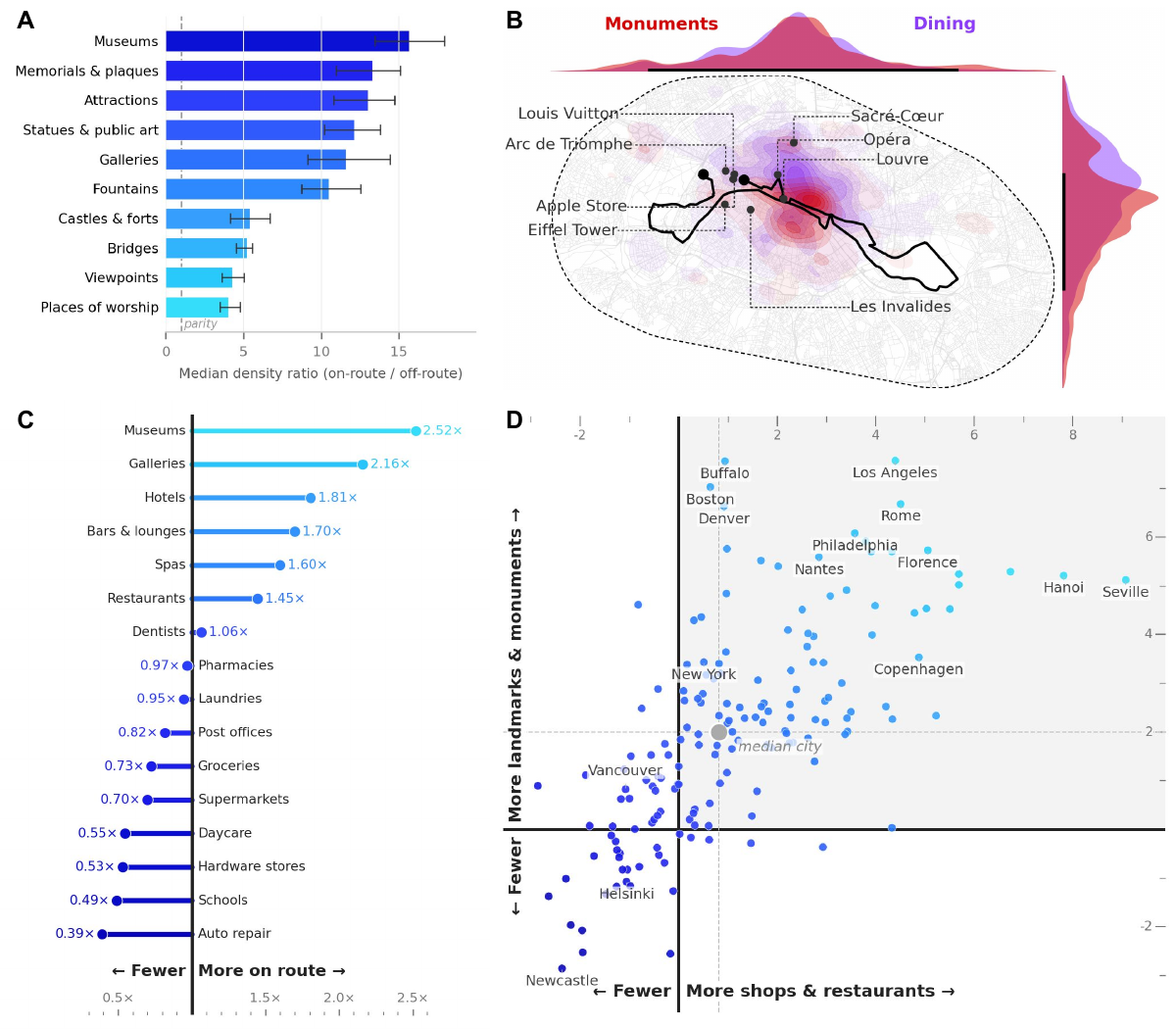}
\caption{\textbf{Marathon biases.} \textbf{A} Considering the density of different monuments and attractions within 200\,m of a marathon route, we see that routes are most skewed towards museums, reflecting centrality, but they are also denser with memorials and plaques, indicating marathons preference historic cores. \textbf{B} Using Paris as an example, we see that that marathon passes through an area with high concentrations of both attractions but also restaurants, although it then leaves these dense areas when it travels through the Bois de Vincennes and the Bois de Boulogne. \textbf{C} Adjusting for the skew towards density, we find that everyday services like schools and daycares are underrepresented but museums and galleries are overrepresented. \textbf{D} We permute each race while fixing start and finish lines and find that race courses are 0.9 and 2 standard deviations from our permutations at the median, for amenities and tourist attractions respectively; relative to these null models, Los Angeles is more biased towards attractions while Seville is more biased towards amenities.}
\label{fig:main-fig1}
\end{figure*}

\section*{Results}
Using our full sample of marathons (see Supplementary Section ~\ref{data}), we begin with a simple test of bias. Within 200\,m of the route, population density is 1.25 times that of the surrounding city and buildings are 1.25 times taller. The area around a race course contains 8.5 times more retail and dining amenities per square kilometre than areas farther afield. Marathons run through dense, central parts of cities. The question is which central parts. (Detailed data showing the results described here can be found in Supplementary Section ~\ref{results}.)

Landmarks and monuments are where the bias is most obvious. As we see in Fig. ~\ref{fig:main-fig1}\textbf{A}, within 200\,m of the route, museums are 15.7 times denser than areas off the route, memorials and plaques 13.3 times, attractions 13.0 times, statues and public art 12.1 times, and galleries 11.6 times. Marathons are efficient ways to see the symbolic city. The strongest landmark and monument bias is in Paris, which show as an example in Fig. ~\ref{fig:main-fig1}\textbf{B} and where the marathon starts and ends at the \emph{Arc de Triomphe} and passes many of the most iconic sites in the city. 

Because routes pass through the densest parts of a city, every amenity class is inflated on-route relative to off-route areas. Yet this inflation is not the same across urban functions. As we show in ~\ref{fig:main-fig1}\textbf{C}, when each category is divided by the route's own average skew, museums are 2.5 times more overrepresented than the city average would suggest, galleries 2.2 times, hotels 1.8 times, bars and lounges 1.7 times, and restaurants 1.5 times. Schools, daycares, hardware stores, and auto repair shops all fall below parity. A within city comparison gives the same result: relative to supermarkets, museums are 3.5 times more overrepresented, galleries 3.2 times, hotels 2.5 times, and restaurants 2.0 times, while schools and auto repair are underrepresented. Across cities, the median route is 2.7 times more skewed toward leisure than toward everyday services. Marathon routes thus do not merely sample dense urban centres; they preference leisure and tourism. We can quantify this effect with principal component analysis, in Supplementary Fig. ~\ref{S5}: a first component explaining 61\% of the variation represents centrality, while a second explaining 8.8\% separates museums, hotels, bars and restaurants from schools, grocers and automotive shops. 

\begin{figure*}[ht!]
\centering
\includegraphics[width=1\textwidth]{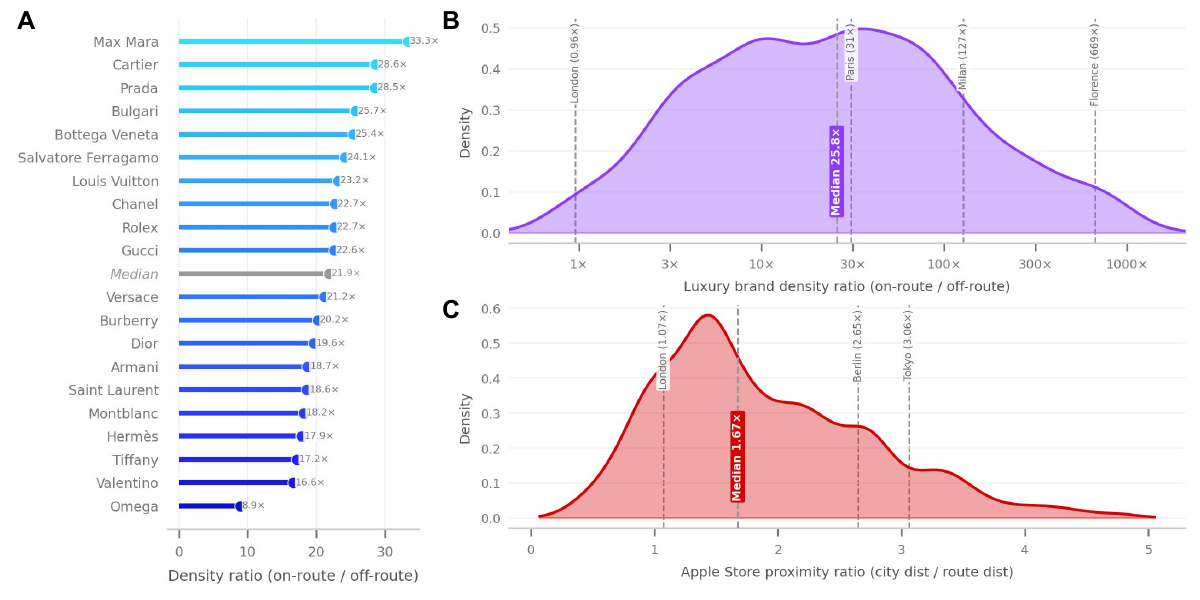}
\caption{\textbf{Luxury brands and marathon routes.} \textbf{A} We create a list of luxury brands and observe density biases on-route and off-route, as above, finding the biases to be far higher than for other classes of amenity. \textbf{B} Considering the distribution of these biases across cities, we see that some marathons---like London's---have no luxury bias at all, while many Italian marathons---like Florence's---have a strong luxury bias. \textbf{C} Apple stores present an interest comparison because the brand makes a conscious effort to locate its stores as ``destinations'' \cite{rao2025toward}, and we see that any given marathon route typically pass closer to an apple store than the average street in that city.}
\label{fig:main-fig2}
\end{figure*}

The commercial landscape is more nuanced, conveying a bias that is subtler than simple consumerism. In 73\% of cities, chain stores represent a smaller share of retail and dining amenities on-route (median 4.5\%) than off-route (median 5.9\%). Marathon routes do not simply trace mass market retail. Yet when we compile a list of luxury brands, the pattern reverses. Excluding 80 cities in our sample without a luxury brand, the median city has about 25.8 times more luxury brands within 200\,m of the route than in the rest of the city, as we see in Fig. ~\ref{fig:main-fig2}\textbf{B}. Three of the top five cities by luxury bias are in Italy. Florence, whose route runs the full length of \emph{Via de' Tornabuoni}, has 44 luxury boutiques on-route and just 3 off-route; Milan has 144 on and 45 off; Rome has 83 on and 33 off. London is an interesting counterexample: its route crosses the city from Greenwich to Green Park but avoids Oxford Street and Regent Street, leaving the luxury ratio near parity.

The same pattern appears when we consider distance instead of density. The average route street is closer than the average city street to most urban features we observe, but again this centrality does not affect all features equally. Route streets are typically about twice as close to tourist attractions like museums as city streets at large, compared with only modest gains for schools or places of worship. After normalising by each race's own average proximity skew, museums remain 1.34 times closer than the route average would imply, attractions 1.29 times, hotels 1.16 times, Apple Stores 1.12 times, and luxury retail 1.11 times; schools and places of worship fall below parity. Apple is an interesting proxy for this study because the company is unusually deliberate about where it places stores, tending toward central and historic sites \cite{rao2025toward}. Shown in Fig. ~\ref{fig:main-fig2}\textbf{C}, in cities with an Apple Store, the average marathon street is 1.68 times closer to one than the average city street.

The remaining question is whether this occurs simply because marathons are, for convenience, in the middle of cities. To assess this we construct a null model that perturbs the route while fixing start and finish lines. A key difference is between circuit races and point to point races: loops permit more bias, because distant start and finish lines reduce freedom. When start and finish are adjacent, the route can be bent around preferred districts; when they are far apart, more of the distance budget is spent connecting the anchors.

Our null model suggests that route bias is not reducible to centrality alone. Relative to alternative routes with the same start and finish, observed marathon courses have higher amenity ratios in 68\% of cities and higher landmark ratios in 82\%. The median observed course is 1.38 times more amenity rich and 1.79 times more landmark rich than the mean null route. The effect is clearer for landmarks: 53\% of races exceed the null at $p < 0.05$, compared with 34\% for amenities, and in about a third of cities the observed route contains more landmarks than every route in the simulated ensemble. Loop races also show more excess over the null than point to point races. The median loop is 1.45 times more amenity rich and 1.85 times more landmark rich than its null mean, compared with 1.12 and 1.42 for point to point races. When we take the z-score of the observed pattern relative to the null distribution in Fig. ~\ref{fig:main-fig1}\textbf{D}, we see that most cities exist in the upper-right quadrant, displaying higher than average amenities and attracts; very few cities---Helsinki and Newcastle are notably exceptions---are in the lower-left quandrant with both fewer amenities and fewer attractions on the actual course relative to the counterfactual routes. The median z-score for observed courses is 0.9 for amenities and 2 for attractions. That the presence of monuments is more sensitive to perturbations makes sense: these routes are intentionally monumental. 

\section*{Discussion}
Marathons are biased on purpose. They are designed to be scenic, logistically tractable, often as flat as possible, and worth traveling for; no one should expect a race course to offer an egalitarian cross section of urban life. This paper has focused on quantifying the biases in marathon routes while generating descriptive statistics that may be of further use. These biases are evident, and while they are also intuitive, they appear to persist beyond reasonable constraints: even after granting planners some fixed parameters, marathon courses still tilt toward tourism, leisure, and prestige. The exquisite nature of route proximity and density to key urban features, that it is so sensitive to adjustment, indicates the benefits of running marathons---that they take you through what may be, according to these data, the most scenic parts of a city (if you look up, of course)---but it also shows its biases. These are not neutral passages through the city, but curated exhibitions of it, deciding which streets stand in for the whole and which parts of ordinary urban life are left out of frame.

\section*{Methods}
We assemble 311 marathon routes from publicly accessible GPX files on Go\&Race. Street networks and landmark tags came from OpenStreetMap via the OSMnx library \cite{boeing2017osmnx}, and amenities, brands, and museums came from Foursquare Open Source Places. For each race we draw a 200\,m corridor around the route and define the surrounding city as the 5\,km buffer around that route with the corridor removed. This gives a local comparison: not the legal city boundary, but the nearby urban fabric the race could plausibly have traversed. (Some marathons, like Boston's, stretch beyond formal city boundaries.) We use the same on-route and off-route comparison for all sources, including amenities, landmarks, buildings, population, and vegetation.

For amenities and landmarks, our main measure is a density ratio. We specify density bias for city $c$ and category $k$ as
\[
R^{\mathrm{dens}}_{ck}
=
\frac{N^{\mathrm{on}}_{ck}/A^{\mathrm{on}}_c}
{N^{\mathrm{off}}_{ck}/A^{\mathrm{off}}_c},
\]
where $N^{\mathrm{on}}_{ck}$ and $N^{\mathrm{off}}_{ck}$ are the numbers of places in category $k$ on- and off-route, and $A^{\mathrm{on}}_c$ and $A^{\mathrm{off}}_c$ are the corresponding areas. Values above one mean that a category is denser near the route than elsewhere in the surrounding city. Because races are naturally concentrated in denser areas, most categories tend to be above parity. To isolate what is unusually concentrated even within the dense central city, we also divide each category's ratio by the geometric mean of all amenity ratios in the same city. For direct within city comparisons, we use supermarkets as a reference. Chain brands are brands that appear in ten or more cities in our sample; luxury brands come from a hand built list matched to the Foursquare brand field.

Counts can be unstable for sparse categories, so we also compute a distance ratio:
\[
R^{\mathrm{prox}}_{ck}
=
\frac{\bar d^{\mathrm{city}}_{ck}}
{\bar d^{\mathrm{route}}_{ck}},
\]
where $\bar d^{\mathrm{route}}_{ck}$ is the mean nearest distance from route streets to the nearest place in category $k$, and $\bar d^{\mathrm{city}}_{ck}$ is the same quantity for streets across the surrounding city. Values above one mean that the route is closer than the city average. We summarize buildings, population, urbanisation, and vegetation with the same on-route and off-route contrast.

To ask whether these patterns are just a product of centrality, we construct a null model on the street network. For each race we keep the observed start and finish, perturb the route incrementally on the network via random walkers, and retain continuous alternative routes that remain close to marathon distance. In short, this method selects four intersections along the race course and moves them with biased random walker, who cannot return to a visited node in the network; each walker moves 20 steps---or 20 intersections. With start and finish lines fixed, we then use Dijkstra's algorithm to join the nodes and reject circuits that are more than 10\% longer or shorter than 42.2\,km. This yields 50 accepted alternatives per race. We then compute the same amenity and landmark ratios for each accepted route and compare the observed course with that null distribution. We report the observed to null ratio and
\[
z=\frac{\mathrm{obs}-\mu_{\mathrm{null}}}{\sigma_{\mathrm{null}}},
\]
where $\mu_{\mathrm{null}}$ and $\sigma_{\mathrm{null}}$ are the mean and standard deviation of the accepted alternatives. Values above zero mean that the observed route is more biased than the average plausible alternative in the same city. We classify races as loop or point-to-point using the gap between start and finish, since distant anchors leave less room for scenic detours.

% \end{linenumbers}

% Bibliography (Nature-style, with DOI support) — style is set in wlscirep.cls
\bibliography{references}

\section*{Acknowledgements}
This research was inspired by the 2026 Barcelona Marathon. 

\section*{Author contributions statement}
\textbf{A.R.} Conceptualization, methodology, analysis, writing.

\subsection*{Data and code availability}
All code used to collect, clean, and analyze the data, and to reproduce the figures, is available at \href{https://github.com/asrenninger/marathons}{a public project repository}. Marathon route geometries were collected from publicly accessible GPX downloads on \href{https://www.goandrace.com/}{Go\&Race}. Street networks and volunteered geographic data were obtained from \href{https://www.openstreetmap.org/}{OpenStreetMap}, and points of interest from \href{https://docs.foursquare.com/data-products/docs/fsq-places-open-source}{Foursquare Open Source Places}. The repository contains cleaned metadata, derived analysis tables, and scripts to rebuild the dataset from the original sources.

\section*{Competing interests}
The authors declare no competing interests.

% =========================
% Embedded Supplementary Information
% =========================
\beginsupplementaryinformation

\begin{center}
{\LARGE Supplementary Information for\\[0.5\baselineskip]
\textbf{The benefits and biases of seeing the word's cities through marathons}\par}
\vspace{0.9\baselineskip}
{\large Andrew Renninger*\par}
\vspace{0.35\baselineskip}
{{\normalsize $^*$Corresponding author: Andrew Renninger (E-mail: andrew.renninger@glasgow.ac.uk)}\par}
\end{center}

% \supplementarytableofcontents
\clearpage

\section{Data}
\label{data}

\begin{figure*}[h!]
\centering
\includegraphics[width=1\textwidth]{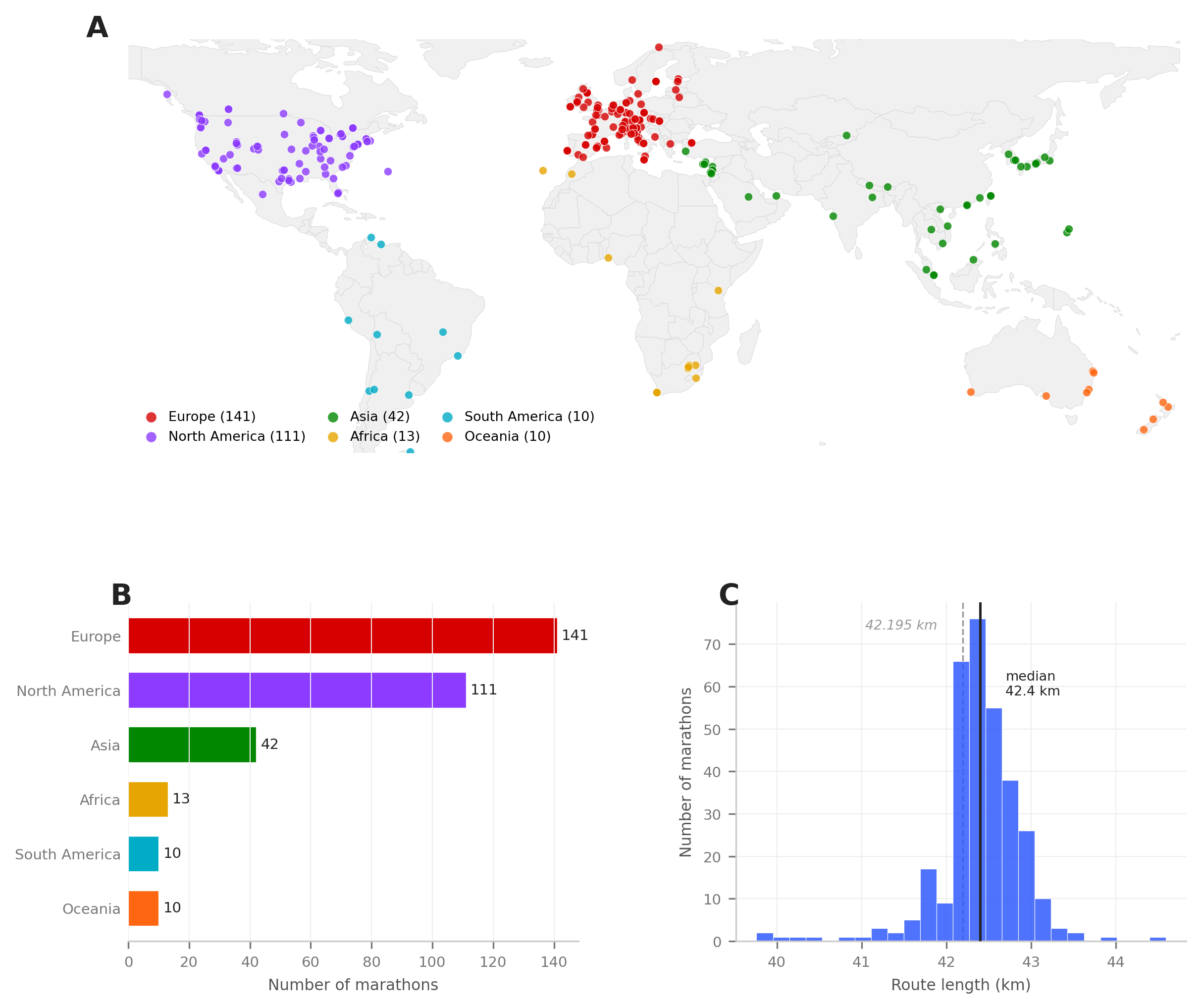}
\caption{\textbf{Data description.} \textbf{A} Locations of the marathons in our sample. The dataset is concentrated in Europe and North America but spans all inhabited continents. \textbf{B} Number of races by continent. Europe and North America account for most of the sample, with smaller samples from Asia, Africa, South America, and Oceania. \textbf{C} Distribution of route lengths derived from the GPX traces. Thankfully, most routes cluster tightly around the official marathon distance of 42.195 km, with a median of 42.4 km. Noise can be explained by imprecision in the GPX traces rather than true errors in the routes.}
\label{S1}
\end{figure*}

\clearpage

\section{Results}
\label{results}

\begin{figure*}[h!]
\centering
\includegraphics[width=1\textwidth]{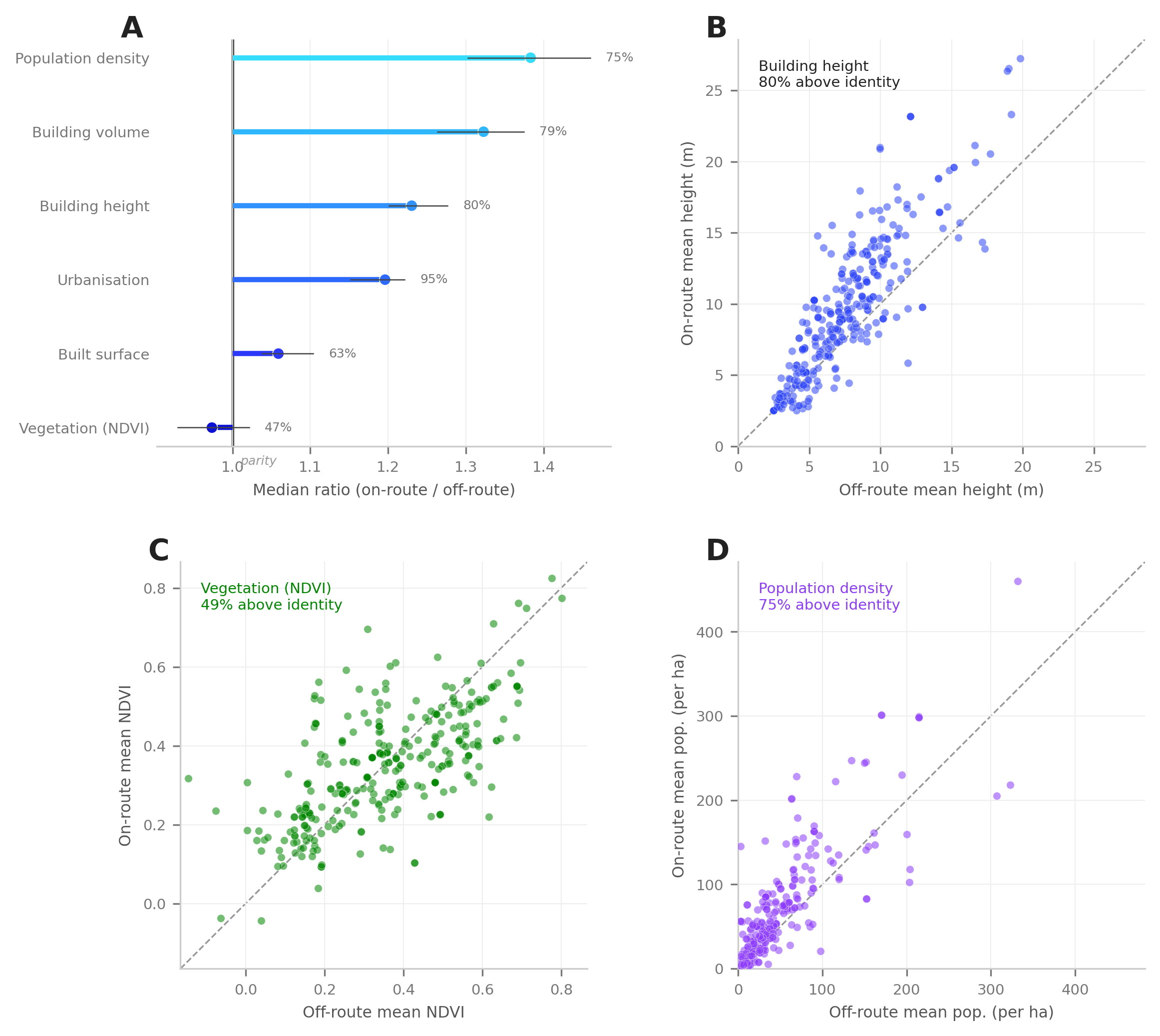}
\caption{\textbf{Population and density.} \textbf{A} Median on route to off route ratios for population density, building volume, building height, urbanisation, built surface, and vegetation (NDVI). Marathon routes tend to pass through denser, taller, and more urbanised parts of cities; vegetation is close to parity. Percent labels give the share of cities above parity. \textbf{B} and \textbf{D} City level comparisons for building height, vegetation, and population density. Most cities lie above identity for building height and population density, while vegetation is much more balanced.}
\label{S2}
\end{figure*}

\clearpage

\begin{figure*}[h!]
\centering
\includegraphics[width=1\textwidth]{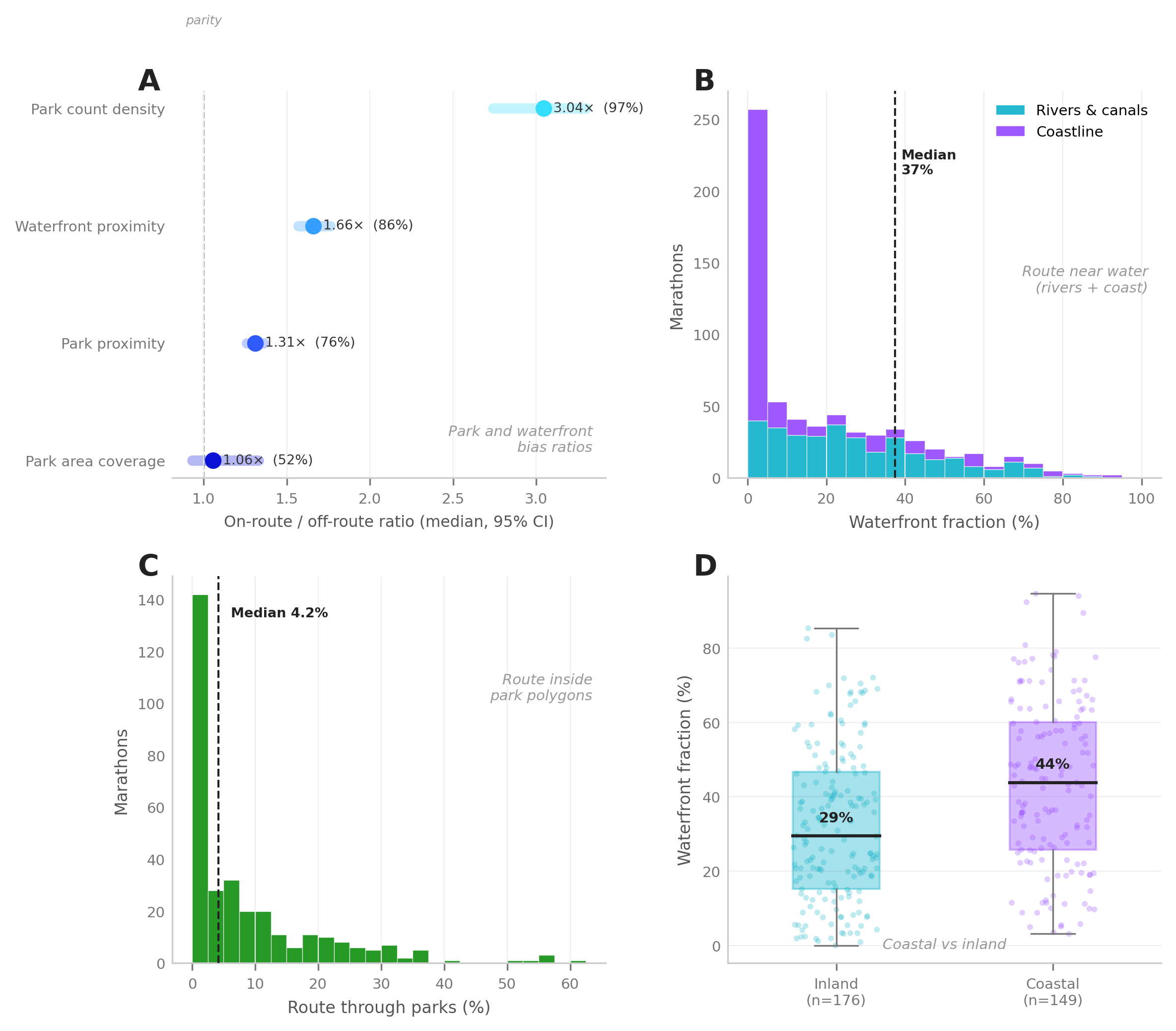}
\caption{\textbf{Parks and water.} \textbf{A} Observed amenity ratios against the mean null ratio for each city. \textbf{B} The same comparison for landmark ratios. Axes are on log scales, and points above the dashed identity line indicate that the observed course is more biased than the average alternative route. This occurs in 68\% of cities for amenities and 82\% for landmarks. \textbf{C} Excess over the null by route type, measured as the observed ratio divided by the mean null ratio. Loop races tend to show more excess than point-to-point races, especially for landmarks; horizontal marks show medians and the dashed line marks parity. \textbf{D} Excess over the null against start finish gap. Routes with larger start finish gaps generally have less room for scenic detours, and landmark excess weakens as this gap grows.}
\label{S3}
\end{figure*}

\clearpage

\begin{figure*}[h!]
\centering
\includegraphics[width=1\textwidth]{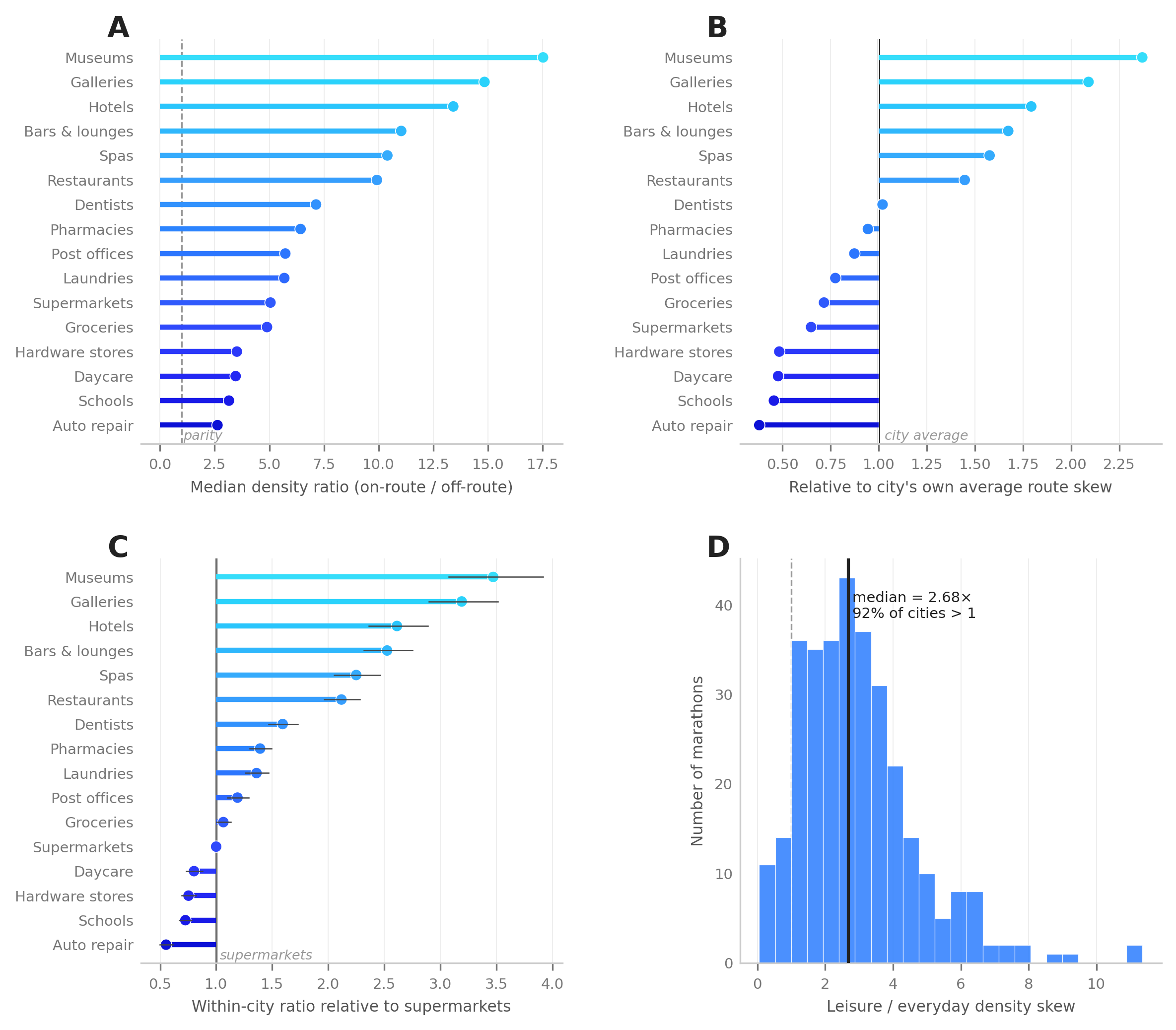}
\caption{\textbf{Density biases.} \textbf{A} Raw on-route to off-route density ratios by amenity type. Nearly every amenity is more common near marathon routes, reflecting centrality, but museums, galleries, hotels, bars and lounges, and restaurants are especially concentrated. \textbf{B} Dividing each category by the route's own average skew isolates which amenities are unusually overrepresented within the dense central city. Museums, galleries, hotels, bars and lounges, and restaurants remain high, while supermarkets, schools, daycares, hardware stores, and auto repair fall below the city average. \textbf{C} Within city paired comparisons relative to supermarkets show the same pattern; points are estimated ratios and horizontal bars are 95\% confidence intervals. \textbf{D} Composite leisure to everyday density skew, where leisure includes restaurants, hotels, galleries, museums, bars and lounges, and spas, and everyday includes supermarkets, groceries, schools, daycares, hardware stores, auto repair, pharmacies, laundries, post offices, and dentists. The median route is 2.68 times more skewed toward leisure than toward everyday services, and 92\% of cities lie above parity.}
\label{S4}
\end{figure*}

\clearpage

\begin{figure*}[h!]
\centering
\includegraphics[width=1\textwidth]{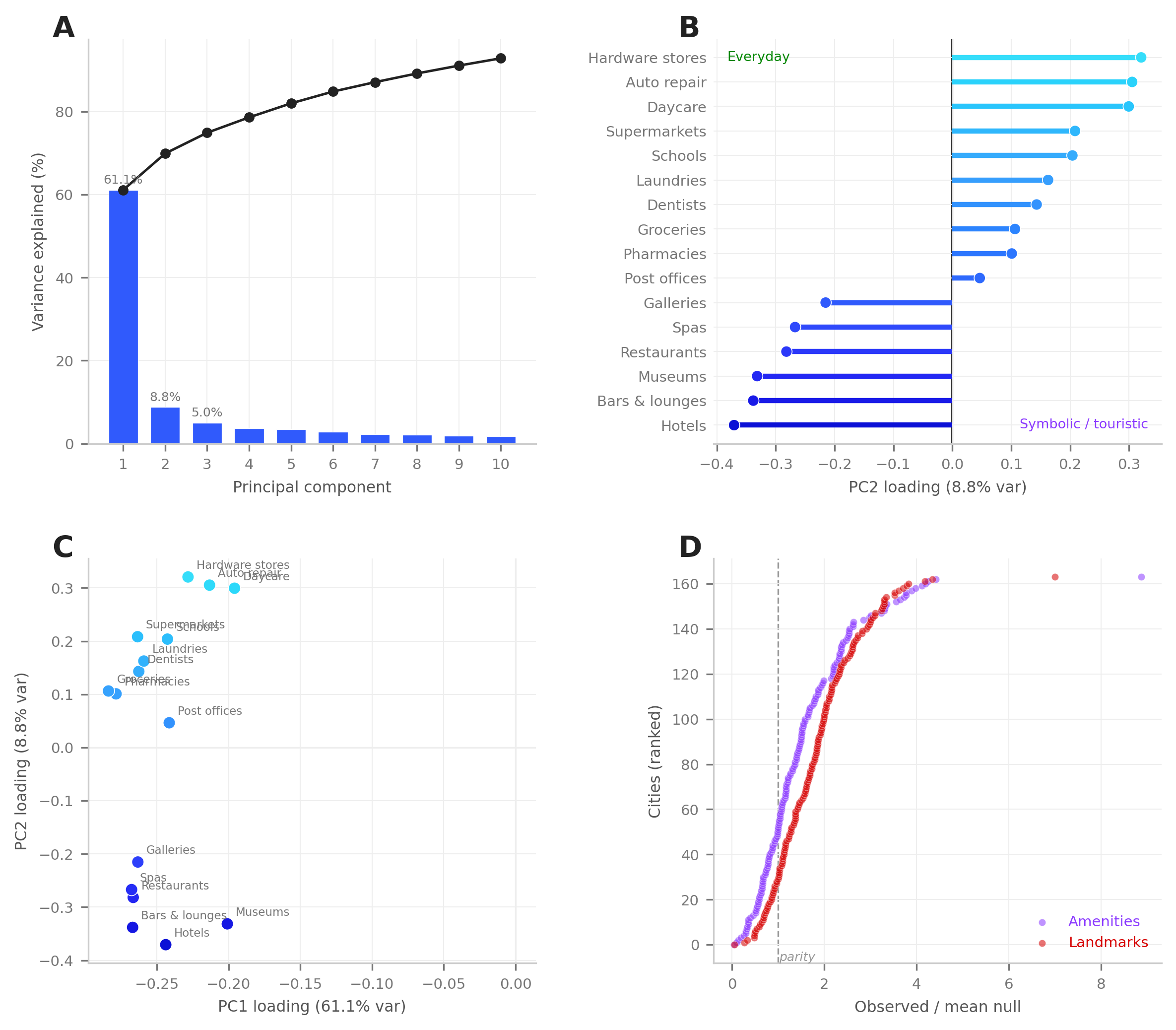}
\caption{\textbf{Residual patterns in route biases.} \textbf{A} Scree plot for principal components estimated from the log density ratios. The first component explains 61.1\% of the variation and mostly captures general centrality. \textbf{B} Loadings on the second component, which explains 8.8\% of the variation. This axis separates everyday services from symbolic and visitor amenities. \textbf{C} Loadings for the first two components. PC1 loads on nearly everything and is best read as dense central city; PC2 separates museums, hotels, bars and lounges, restaurants, and spas from schools, supermarkets, hardware stores, and auto repair. \textbf{D} Ranked city level excess over the null, measured as the observed ratio divided by the mean null ratio. Most cities lie above parity on both amenities and landmarks, but the excess is stronger and more widespread for landmarks.}
\label{S5}
\end{figure*}

\clearpage

\begin{figure*}[h!]
\centering
\includegraphics[width=1\textwidth]{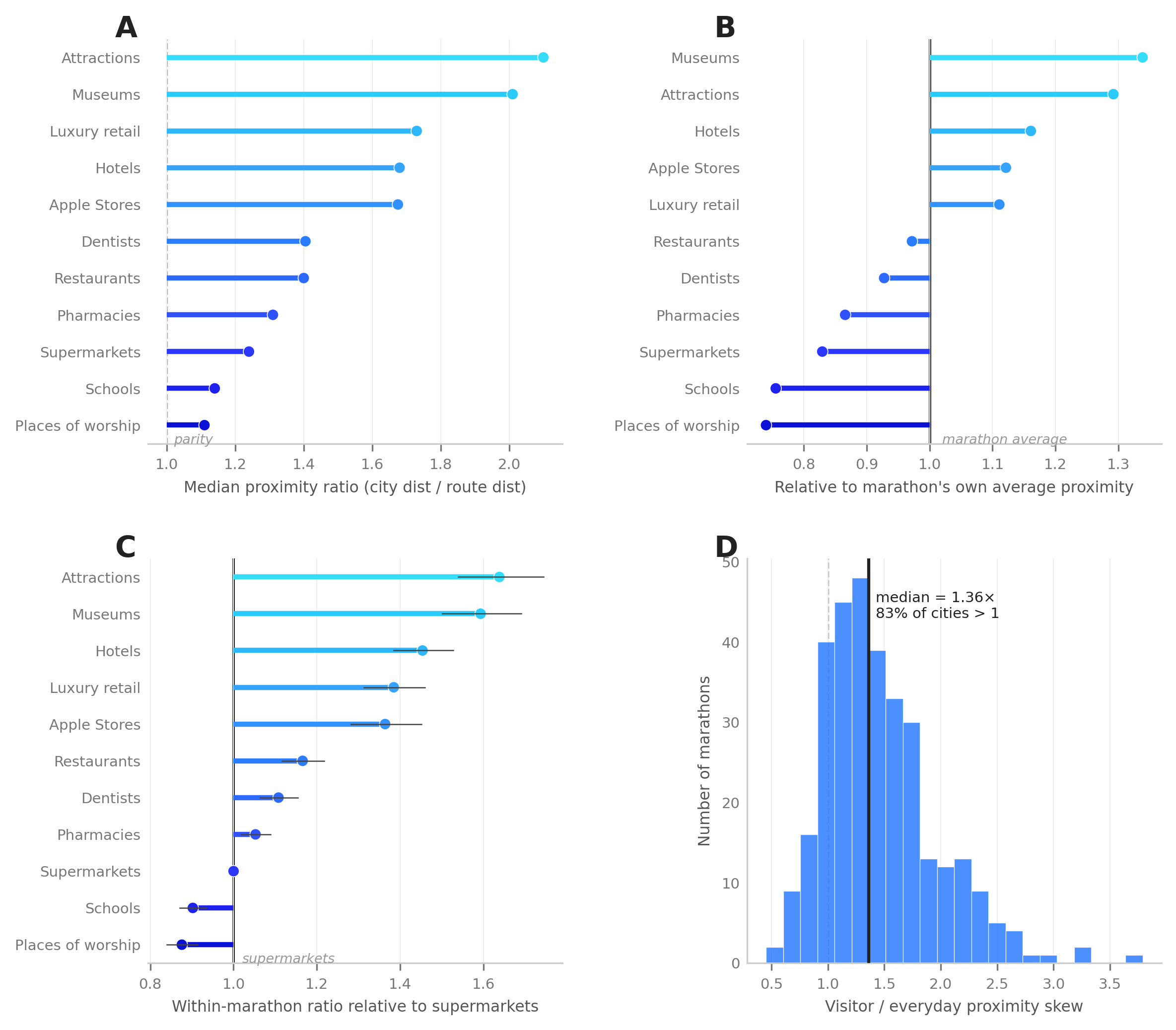}
\caption{\textbf{Distance biases.} \textbf{A} Median proximity ratios, defined as mean city distance divided by mean route distance. Values above one mean that the route is closer than the city average. Route streets are especially close to attractions, museums, luxury retail, hotels, and Apple Stores. \textbf{B} Dividing each category by the marathon's own average proximity skew shows that museums, attractions, hotels, Apple Stores, and luxury retail remain unusually close even after accounting for general centrality. Schools and places of worship fall below the marathon average. \textbf{C} Within marathon paired comparisons relative to supermarkets show the same pattern; points are estimated ratios and horizontal bars are 95\% confidence intervals. \textbf{D} Composite visitor to everyday proximity skew, where visitor space includes museums, attractions, hotels, luxury retail, Apple Stores, and restaurants, and everyday space includes supermarkets, pharmacies, schools, and dentists. The median route is 1.36 times more skewed toward visitor facing space, and 83\% of cities lie above parity.}
\label{S6}
\end{figure*}

\clearpage

\begin{figure*}[h!]
\centering
\includegraphics[width=1\textwidth]{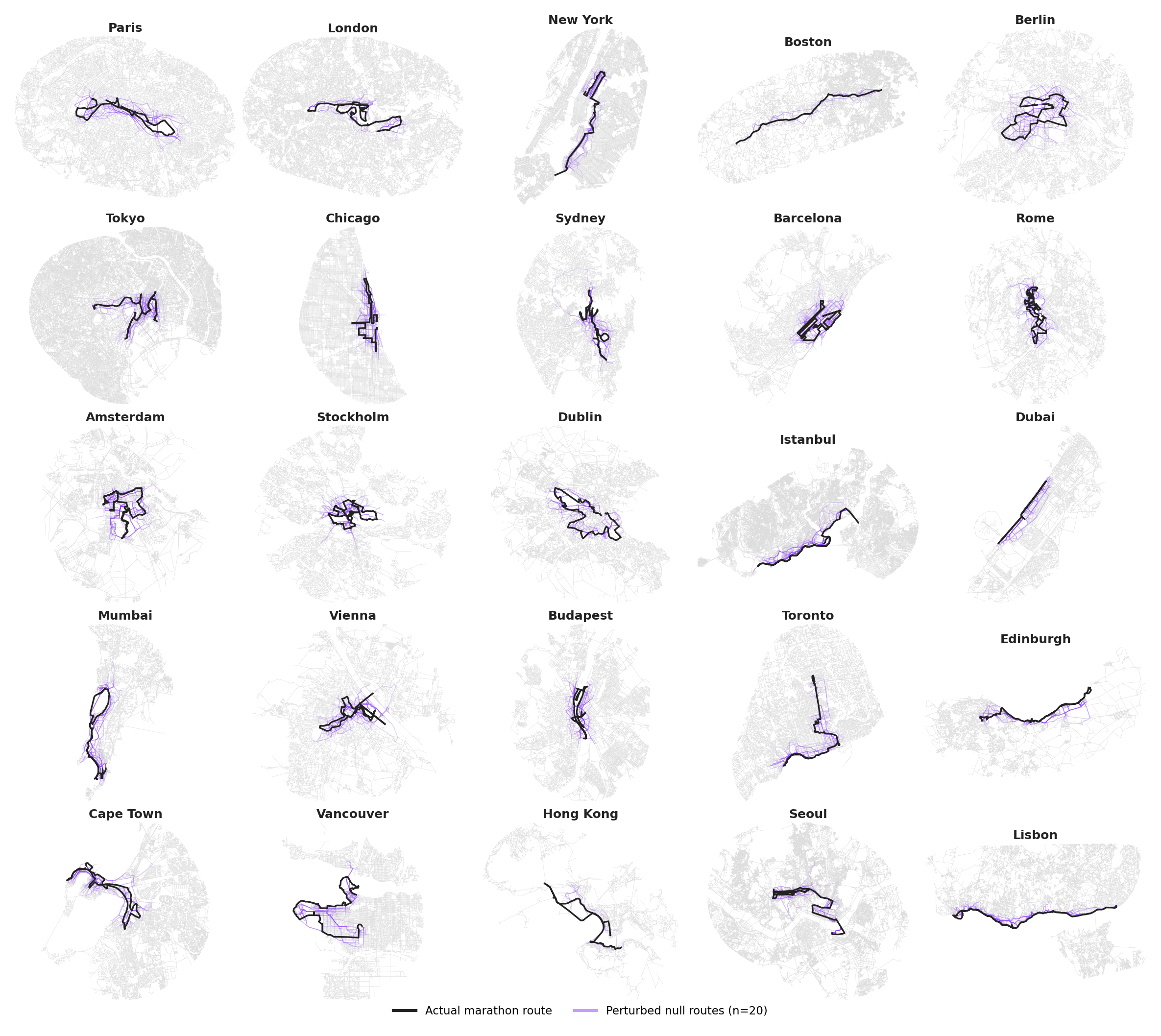}
\caption{\textbf{Null model examples.} Black lines show observed marathon routes and purple lines show twenty alternative $\sim$42 km routes generated on the same street network. The examples span both loop and point-to-point races and a wide range of urban forms. The goal of this null is achieve a balance between realism and possibility, to show plausible alternative ways of spending marathon distance in the same approximate areas while respecting broad constraints.}
\label{S7}
\end{figure*}

\clearpage

\begin{figure*}[h!]
\centering
\includegraphics[width=1\textwidth]{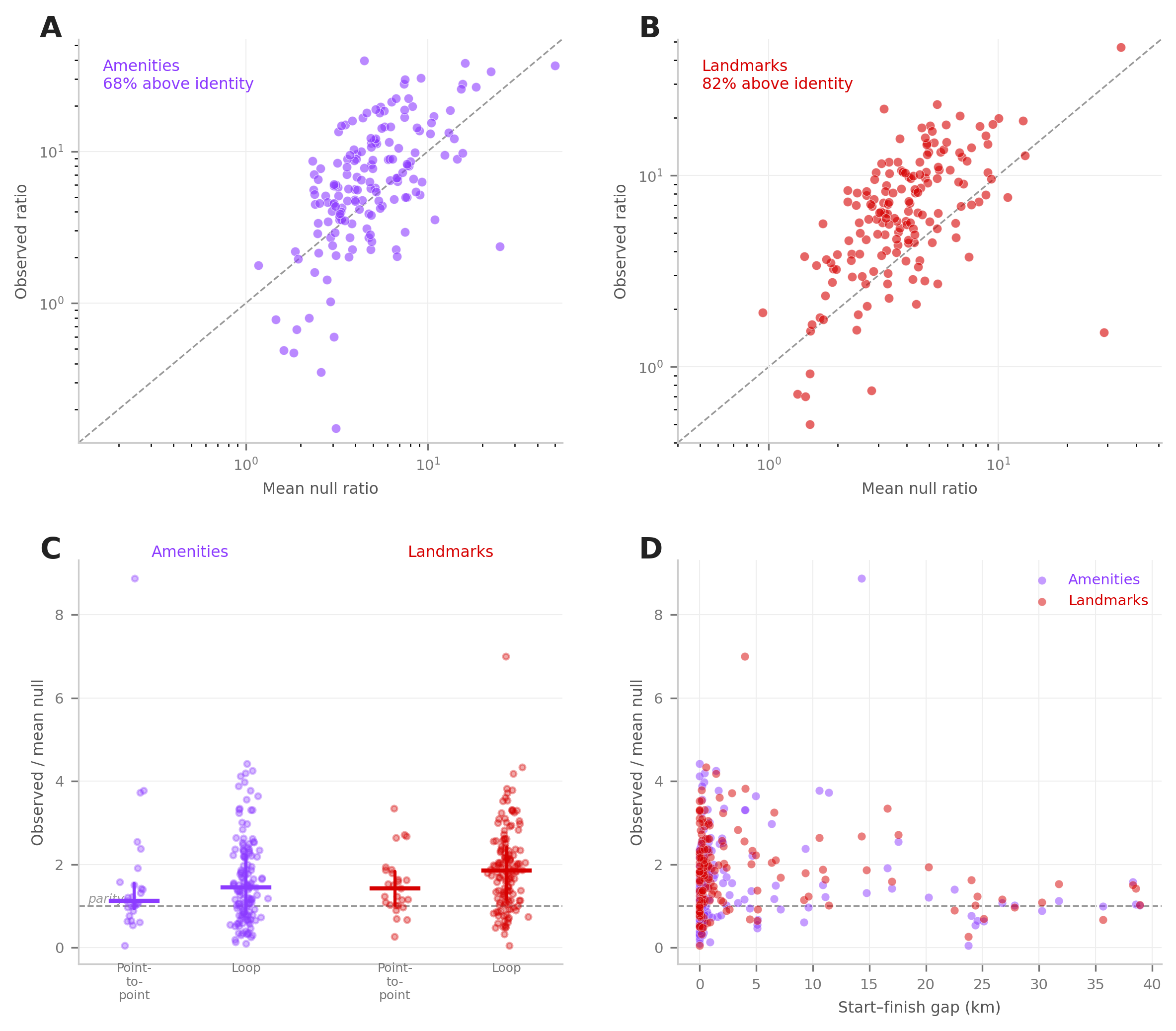}
\caption{\textbf{Null model results.} \textbf{A} Observed amenity ratios against the mean null ratio for each city. \textbf{B} The same comparison for landmark ratios. Axes are on log scales, and points above the dashed identity line indicate that the observed course is more biased than the average alternative route. This occurs in 68\% of cities for amenities and 82\% for landmarks. \textbf{C} Excess over the null by route type, measured as the observed ratio divided by the mean null ratio. Loop races tend to show more excess than point-to-point races, especially for landmarks; horizontal marks show medians and the dashed line marks parity. \textbf{D} Excess over the null against start finish gap. Routes with larger start finish gaps generally have less room for scenic detours, and landmark excess weakens as this gap grows.}
\label{S8}
\end{figure*}

\clearpage

% \begin{table}[h!]
% \centering
% \caption{\textbf{Supplementary Table title.} Example table scaffold.}
% \label{T1}
% \begin{tabular}{lcc}
% \toprule
% Item & Value A & Value B\\
% \midrule
% Thing 1 & 1.23 & 4.56\\
% Thing 2 & 7.89 & 0.12\\
% \bottomrule
% \end{tabular}
% \end{table}

% \clearpage

\section{Methods}

\subsection*{Route sample and spatial data}
We download publicly accessible GPX files for marathon routes from Go\&Race and retain one route geometry per race after removing duplicates and files with incomplete traces. We project all geometries to a metric coordinate system before buffering, area calculations, and distance calculations. We end up with a sample of 311 marathons; some figures may use different counts where a category is missing or a null ensemble could not be completed. Street networks and volunteered geographic data came from OpenStreetMap via the OSMnx library \cite{boeing2017osmnx}. Amenities, brands, and museums came from Foursquare Open Source Places.

\subsection*{On route and off route space}
For each race we drew a 200\,m corridor around the route line and treated this as on route space. To define the surrounding city, we drew a 5\,km buffer around the route and removed the 200\,m corridor. The corridor is therefore the on-route area, and the remainder of the 5\,km buffer is the off-route area. Areas were measured in square kilometres.

This comparison is simple by design. It does not attempt to define the legal city boundary or the full built up extent of each city, and it accommodates the fact that some marathons stretch beyond formal city limits---as in Boston. Instead, it compares the corridor actually used by the race with the nearby urban fabric that the race could plausibly have traversed.

\subsection*{Density ratios}
For each city $c$ and amenity class $k$, we count the number of places in the route corridor and in the broader area. These counts were converted to densities by dividing by area. Our main density ratio is

\[
R^{\mathrm{dens}}_{ck}
=
\frac{N^{\mathrm{on}}_{ck}/A^{\mathrm{on}}_c}
{N^{\mathrm{off}}_{ck}/A^{\mathrm{off}}_c},
\]

where $N^{\mathrm{on}}_{ck}$ and $N^{\mathrm{off}}_{ck}$ are the counts of category $k$ on and off route, and $A^{\mathrm{on}}_c$ and $A^{\mathrm{off}}_c$ are the corresponding areas. Values above one mean that the category is denser near the race course than elsewhere in the surrounding city.

Because marathon routes usually pass through dense central parts of cities, almost every category tends to be above parity. To separate this general centrality from more selective route design, we also divide each category's density ratio by the geometric mean of all amenity ratios in the same city. A value above one on this measure means that a category is more concentrated than the route's overall centrality would predict.

To compare categories directly within the same city, we use supermarkets as a reference point. For each category we computed the paired log difference

\[
\log R^{\mathrm{dens}}_{ck} - \log R^{\mathrm{dens}}_{c,\mathrm{supermarket}},
\]

averaged that quantity across cities, and exponentiated to produce a ratio. The confidence intervals come from standard errors of these paired log differences.

\subsection*{Distance ratios}
Counts and densities can be unstable for sparse categories, so we also compute a distance measure. For each city and category, we calculate the mean nearest distance from route street locations to the nearest place in that category, and the corresponding mean nearest distance from street locations across the full city buffer. Our proximity ratio is

\[
R^{\mathrm{prox}}_{ck}
=
\frac{\bar d^{\mathrm{city}}_{ck}}
{\bar d^{\mathrm{route}}_{ck}}.
\]

Values above one mean that the average route street is closer to that category than the average city street.

As with the density analysis, we also divide each category's proximity ratio by the geometric mean of all proximity ratios in the same city. This tells us whether a category is unusually close even after accounting for the general fact that marathon routes tend to run through central areas. We again use supermarkets as a reference point for within city paired comparisons.

\subsection*{Composite measures}
To summarise the contrast between visitor space and everyday urban space, we constructed simple composites using geometric means across related categories.

For the density analysis, the leisure composite used restaurants, hotels, galleries, museums, bars and lounges, and spas. The everyday composite use supermarkets, groceries, schools, daycares, hardware stores, automotive shops, pharmacies, laundries, post offices, and dentists. The leisure to everyday skew is the ratio of those two geometric means.

For the proximity analysis, the visitor composite used museums, attractions, hotels, luxury retail, Apple Stores, and restaurants. The everyday composite used supermarkets, pharmacies, schools, and dentists. Again, the reported skew is the ratio of the two geometric means.

These composites are descriptive. Their purpose is not to define a universal boundary between tourist and ordinary life, but to summarise a pattern that is already visible in the category by category results.

\subsection*{Brands, chains, and luxury retail}
Chain share was measured among retail and dining amenities only. A brand is treated as a chain if it appears in ten or more cities in our sample. This yields a conservative measure of large, widely distributed brands.

We identify luxury retail by asking a large language model decide which brands in the Foursquare data are ``luxury'', beginning with a seed that uses the constituent companies from luxury conglomerates like LVMH, Kering, Richmont, and Hermès to guide the model. We then vetted these brands manually. The list is designed to capture globally legible premium fashion and jewellery brands rather than all expensive commerce. Apple Stores were treated separately.

\subsection*{Built environment and population}
Using data from the Global Human Settlement Layer (GHSL) and Sentinel-2, we also summarise several descriptive features of the built environment around the route, including population density, building height, building volume, building surface, urbanisation, and vegetation. For each measure we compare the mean value within the 200\,m corridor to the mean value in the surrounding off route area. These quantities are used as simple route diagnostics rather than as inputs to the main route bias measures.

\subsection*{Principal component analysis}
We use principal component analysis only as a descriptive check on the amenity density ratios. The input matrix has one row per marathon and one column per amenity ratio. We take logs of the ratios, replace missing values with the median of the corresponding column, and standardise each column before running the analysis.

The first component captures the broad fact that some routes run through more central and more amenity dense space than others. In our data this component explains 61.1\% of the variation. The second component captures something different: not how central a route is, but what kind of central city it selects. In our data, museums, hotels, bars and lounges, restaurants, and spas load on one side of this axis, while schools, supermarkets, hardware stores, daycares, and auto repair load on the other. This second component explains 8.8\% of the variation. The signs of principal components are arbitrary, but the splits are informative.

\subsection*{Null routes}
The main descriptive comparison is between the observed route and the surrounding city. To ask whether the observed pattern could be explained by route geometry alone, we also generate alternative marathon routes on the same street network while keeping the observed start and finish fixed and keeping total length close to marathon distance. We create these alternatives by incrementally perturbing the observed course on the network using random walkers. We sample 1 in 4 intersections from the race and, for each intersection, a random walker moves 20 steps at random across the street network without return to previously occupied location. We then rejoin the intersections using Dijkstra's algorithm and reject circuits that are deviate by 10\% from an the official marathon length.

We repeat this process such that for each race we retain 50 accepted alternatives. We then compute the same amenity and landmark ratios for each accepted alternative as for the observed route. This yields a null distribution for every race, summarised by its mean, median, quartiles, standard deviation, and maximum. We compared the observed route with this distribution using the ratio of the observed value to the null mean, an empirical tail probability, and the z score

\[
z = \frac{\mathrm{obs} - \mu_{\mathrm{null}}}{\sigma_{\mathrm{null}}}.
\]

where values above zero indicate that the observed route is more biased than the average alternative route in the same city.

\clearpage

\end{document}